\documentclass[a4paper,10pt]{article}
\usepackage{graphicx}

\newcommand{\CTT}{C^{\rm TT}_\ell}
\newcommand{\CTE}{C^{\rm TE}_\ell}
\newcommand{\CEE}{C^{\rm EE}_\ell}
\newcommand{\CBB}{C^{\rm BB}_\ell}
\newcommand{\CTB}{C^{\rm TB}_\ell}
\newcommand{\CEB}{C^{\rm EB}_\ell}
\newcommand{\EI}{{\cal E}_{\rm Inf}}

\newcommand{\mpc}{\ensuremath{{\rm\,Mpc}}}

\newcommand{\hmpc}{\ensuremath{h^{-1}{\rm\,Mpc}}}

\newcommand{\Omhh}{\Omega_mh^2}
\newcommand{\Obhh}{\Omega_bh^2}

\def\lsim{\mathrel{\rlap{\lower4pt\hbox{\hskip1pt$\sim$}}
    \raise1pt\hbox{$<$}}}                
\def\gsim{\mathrel{\rlap{\lower4pt\hbox{\hskip1pt$\sim$}}
    \raise1pt\hbox{$>$}}}                

\def\be{\begin{equation}}
\def\ee{\end{equation}}
\def\bea{\begin{eqnarray}}
\def\eea{\end{eqnarray}}

\def\biposh#1#2#3{ A^{#2}_{#3}}

\begin{document}


\title{`Standard' Cosmological model \& beyond with CMB}

\author{Tarun Souradeep \\Inter-University Centre for Astronomy and
Astrophysics,\\ Post Bag 4, Ganeshkhind, Pune 411~007, India.\\
email: tarun@iucaa.ernet.in}

\maketitle

\begin{abstract}
Observational Cosmology has indeed made very rapid progress in the
past decade. The ability to quantify the universe has largely improved
due to observational constraints coming from structure formation
Measurements of CMB anisotropy and, more recently, polarization have
played a very important role. Besides precise determination of various
parameters of the `standard' cosmological model, observations have
also established some important basic tenets that underlie models of
cosmology and structure formation in the universe -- `acausally'
correlated initial perturbations in a flat, statistically isotropic
universe, adiabatic nature of primordial density perturbations.  These
are consistent with the expectation of the paradigm of inflation and
the generic prediction of the simplest realization of inflationary
scenario in the early universe. Further, gravitational instability is
the established mechanism for structure formation from these initial
perturbations. The signature of primordial perturbations observed as
the CMB anisotropy and polarization is the most compelling evidence
for new, possibly fundamental, physics in the early universe. The
community is now looking beyond the estimation of parameters of a
working `standard' model of cosmology for subtle, characteristic
signatures from early universe physics.
\end{abstract}

\section{Introduction}	
 
The `standard' model of cosmology must not only explain the dynamics
of the homogeneous background universe, but also satisfactorily
describe the perturbed universe -- the generation, evolution and
finally, the formation of large scale structures in the universe. It
is fair to say much of the recent progress in cosmology has come from
the interplay between refinement of the theories of structure
formation and the improvement of the observations.

The transition to precision cosmology has been spearheaded by
measurements of CMB anisotropy and, more recently,
polarization. Despite its remarkable success, the `standard' model of
cosmology remains largely tied to a number of fundamental assumptions
that have yet to find complete and precise observational
verification~: the Cosmological Principle, the paradigm of inflation
in the early universe and its observable consequences (flat spatial
geometry, scale invariant spectrum of primordial seed perturbations,
cosmic gravitational radiation background etc.).  Our understanding of
cosmology and structure formation necessarily depends on the rather
inaccessible physics of the early universe that provides the stage for
scenarios of inflation (or related alternatives).  The CMB anisotropy
and polarization contains information about the hypothesized nature of
random primordial/initial metric perturbations -- (Gaussian)
statistics, (nearly scale invariant) power spectrum, (largely)
adiabatic vs.  iso-curvature and (largely) scalar vs. tensor
component.  The `default' settings in brackets are motivated by
inflation.  The signature of primordial perturbations on super-horizon
scales at decoupling in the CMB anisotropy and polarization are the
most definite evidence for new physics (eg., inflation ) in the early
universe that needs to be uncovered. However, the precision estimation
of cosmological parameters implicitly depend on the assumed form of
the initial conditions such as the primordial power spectrum, or,
explicitly on the scenario of generation of initial
perturbations~\cite{recons_us,shaf_sour09}.

Besides precise determination of various parameters of the `standard'
cosmological model, observations have also begum to establish (or
observationally query) some of the important basic tenets of cosmology
and structure formation in the universe -- `acausally' correlated
initial perturbations, adiabatic nature of primordial density
perturbations, gravitational instability as the mechanism for
structure formation. We have inferred a spatially flat universe where
structures form by the gravitational evolution of nearly scale
invariant, adiabatic perturbations in the non--baryonic cold dark
matter. There is a dominant component of dark energy that does not
cluster (on astrophysical scales).  We briefly review the observables
from the CMB sky and importance to understanding cosmology in
section~\ref{cmb} Most recent estimates of the cosmological parameters
are available and best obtained from recent literature,
eg. Ref.\cite{lar_wmap10} and, hence, is not given in the article.
The main theme of the article is to highlight \footnote{The article
does {\em not} attempt at a review and is far from being exhaustive in
the coverage of the science and literature.} the success of recent
cosmological observations in establishing some of the fundamental
tenets of cosmology and structure~:

\begin{itemize}

\item{} Statistical Isotropy of the universe (Sec.~\ref{SI});

\item{} Gravitational instability mechanism for structure
formation(Sec.~\ref{GI});

\item{} Primordial perturbations from Inflation.(Sec.~\ref{PI}).

\end{itemize}

Up to this time, the attention of the community has been largely focused on
estimating the cosmological parameters.  The next decade would see
increasing efforts to observationally test fundamental tenets of the
cosmological model and search for subtle deviations from the same
using the CMB anisotropy and polarization measurements and related LSS
observations, galaxy survey, gravitational lensing, etc.

\section{CMB observations and cosmological parameters}
\label{cmb}

The angular power spectra of the Cosmic Microwave Background
temperature fluctuations ($C_\ell$)have become invaluable observables
for constraining cosmological models. The position and amplitude of
the peaks and dips of the $C_\ell$ are sensitive to important
cosmological parameters, such as, the relative density of matter,
$\Omega_0$; cosmological constant, $\Omega_\Lambda$; baryon content,
$\Omega_B$; Hubble constant, $H_0$ and deviation from flatness
(curvature), $\Omega_K$.

\begin{figure}
\begin{center}
\includegraphics[scale= 0.4]{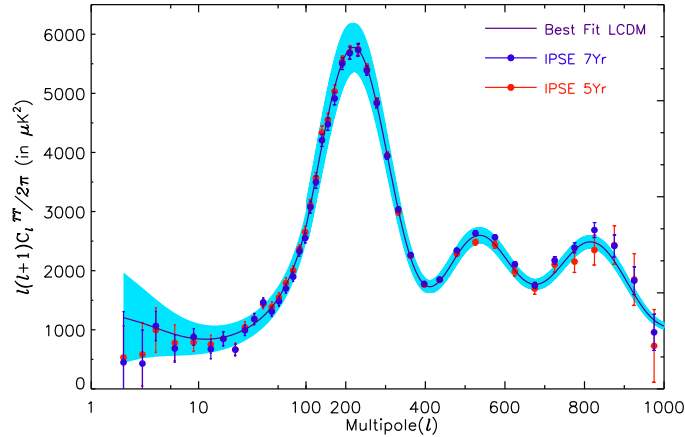}
\caption{ The angular power spectrum estimated from the
multi-frequency five and seven year WMAP data. The result from IPSE a
self-contained model free approach to foreground removal
~\protect{\cite{samal10sah06}} matches that obtained by the WMAP
team. The solid curve showing prediction of the best fit power-law,
flat, $\Lambda$CDM model threads the data points closely.[Fig. courtesy: Tuhin Ghosh] }
\label{WMAPCL}
\end{center}
\end{figure}

  The angular spectrum of CMB temperature fluctuations has been
measured with high precision on up to angular scales ($\ell \sim
1000$) by the WMAP experiment \cite{lar_wmap10}, while smaller angular
scales have been probed by ground and balloon-based CMB experiments
such as ACBAR, QuaD and ACT~\cite{quad09,acbar09act10}.  These data
are largely consistent with a $\Lambda$CDM model in which the Universe
is spatially flat and is composed of radiation, baryons, neutrinos
and, the exotic, cold dark matter and dark energy. The exquisite
measurements by the Wilkinson Microwave Anisotropy Probe (WMAP) mark a
successful decade of exciting CMB anisotropy measurements and are
considered a milestone because they combine high angular resolution
with full sky coverage and extremely stable ambient condition (that
control systematics) allowed by a space mission .  Figure~\ref{WMAPCL}
shows the angular power spectrum of CMB temperature fluctuations
obtained from the 5 \& 7-year WMAP data~\cite{samal10sah06}.

The measurements of the anisotropy in the cosmic microwave background
(CMB) over the past decade has led to `precision cosmology'.
Observations of the large scale structure in the distribution of
galaxies, high redshift supernova, and more recently, CMB
polarization, have provided the required complementary
information. The current up to date status of cosmological parameter
estimates from joint analysis of CMB anisotropy and Large scale
structure (LSS) data is usually best to look up in the parameter
estimation paper accompanying the most recent results announcement of
a major experiment, such as recent WMAP release~\cite{lar_wmap10}.

One of the firm predictions of this working `standard' cosmological
model is linear polarization pattern ($Q$ and $U$ Stokes parameters)
imprinted on the CMB at last scattering surface. Thomson scattering
generates CMB polarization anisotropy at decoupling~\cite{cmb_polar}.
This arises from the polarization dependence of the differential cross
section: $d\sigma/d\Omega\propto |\epsilon'\cdot\epsilon|^2$, where
$\epsilon$ and $\epsilon'$ are the incoming and outgoing polarization
states~\cite{rad_basics} involving linear polarization only.  A local
quadrupole temperature anisotropy produces a net polarization, because
of the $\cos^2\theta$ dependence of the cross section.  A net pattern
of linear polarization is retained due to local quadrupole intensity
anisotropy of the CMB radiation impinging on the electrons at
$z_{rec}$. The polarization pattern on the sky can be decomposed in
the two kinds with different parities.  The even parity pattern arises
as the gradient of a scalar field called the $E$--mode.  The odd
parity pattern arises from the `curl' of a pseudo-scalar field called
the $B$--mode of polarization.  Hence the CMB sky maps are
characterized by a triplet of random scalar fields: $X(\hat{n})\equiv
\{T(\hat{n})$, $E(\hat{n})$, $B(\hat{n})\}$.  For Gaussian CMB sky,
there are a total of 4 power spectra that characterize the CMB
signal~: $\CTT, \CTE, \CEE, \CBB$. Parity conservation eliminates the
two other possible power spectra, $\CTB$ \& $\CEB$. While CMB
temperature anisotropy can also be generated during the propagation of
the radiation from the last scattering surface, the CMB polarization
signal can be generated primarily at the last scattering surface,
where the optical depth transits from large to small values. The
polarization information complements the CMB temperature anisotropy by
isolating the effect at the last scattering surface from effects along
the line of sight.

The CMB polarization is an even cleaner probe of early universe
scenarios that promises to complement the remarkable successes of CMB
anisotropy measurements.  The CMB polarization signal is much smaller
than the anisotropy signal. Measurements of polarization at
sensitivities of $\mu K$ (E-mode) to tens of $nK$ level (B-mode) pose
stiff challenges for ongoing and future experiments.

 After the first detection of CMB polarization spectrum by the Degree
Angular Scale Interferometer (DASI) on the intermediate band of
angular scales ($l\sim 200-440$) in late 2002~\cite{kov_dasi02}, the
field has rapidly grown, with measurements coming in from a host of
ground--based and balloon--borne dedicated CMB polarization
experiments.  The full sky E-mode polarization maps and polarization
spectra from WMAP were a new milestone in CMB
research~\cite{pag_wmap06,kog_wmap03}. The most current CMB
polarization measurement of $\CTT$, $\CTE$ and $\CEE$ and a
non--detection of $B$--modes come from QUaD and BICEP. They also
report interesting upper limits $\CTB$ or $\CEB$, over and above
observational artifacts~\cite{parviol_quad}. A non-zero detection of
$\CTB$ or $\CEB$, over and above observational artifacts, could be
tell-tale signatures of exotic parity violating physics~\cite{parviol}
and the CMB measurements put interesting limits on these possibilities.

While there has been no detection of cosmological signal in B-mode of
polarization, the lack of $B$--mode power suggests that foreground
contamination is at a manageable level which is good news for future
measurements.  The Planck satellite launched in May 2009 will greatly
advance our knowledge of CMB polarization by providing
foreground/cosmic variance--limited measurements of $\CTE$ and $\CEE$
out beyond $l\sim 1000$.  We also expect to detect the weak lensing
signal, although with relatively low precision. Perhaps, Planck could
detect inflationary gravitational waves if they exist at a level of
$r\sim 0.1$. In the future, a dedicated CMB polarization mission is
under study at both NASA and ESA in the time frame 2020+.  These
primarily target the $B$-mode polarization signature of gravity waves,
and consequently, identify the viable sectors in the space of
inflationary parameters.

\section{Statistical Isotropy of the universe}
\label{SI}

The {\em Cosmological Principle} that led to the idealized FRW
universe found its strongest support in the discovery of the (nearly)
isotropic, Planckian, Cosmic Microwave Background. The isotropy around
every observer leads to spatially homogeneous cosmological models.
The large scale structure in the distribution of matter in the
universe (LSS) implies that the symmetries incorporated in FRW
cosmological models ought to be interpreted statistically.

The CMB anisotropy and its polarization is currently the most
promising observational probe of the global spatial structure of the
universe on length scales close to, and even somewhat beyond, the
`horizon' scale ($\sim c H_0^{-1}$).  The exquisite measurement of the
temperature fluctuations in the CMB provide an excellent test bed for
establishing the statistical isotropy (SI) and homogeneity of the
universe. In `standard' cosmology, CMB anisotropy signal is expected
to be statistically isotropic, i.e., statistical expectation values of
the temperature fluctuations $\Delta T(\hat q)$ are preserved under
rotations of the sky. In particular, the angular correlation function
$C(\hat{q},\, \hat{q}^\prime)\equiv\langle\Delta T(\hat q)\Delta
T(\hat q^\prime)\rangle$ is rotationally invariant for Gaussian
fields. In spherical harmonic space, where $\Delta T(\hat q)=
\sum_{lm}a_{lm} Y_{lm}(\hat q)$, the condition of {\em statistical
isotropy} (SI) translates to a diagonal $\langle a_{lm} a^*_{l^\prime
m^\prime}\rangle=C_{l} \delta_{ll^\prime}\delta_{mm^\prime}$ where
$C_l$, is the widely used angular power spectrum of CMB anisotropy.
The $C_l$ is a complete description only of (Gaussian) SI CMB sky CMB
anisotropy and  would be (in principle) an inadequate measure for
comparing models when SI is violated~\cite{bps}.

Interestingly enough, the statistical isotropy of CMB has come under a
lot of scrutiny after the WMAP results. Tantalizing evidence of SI
breakdown (albeit, in very different guises) has mounted in the {\it
WMAP} first year sky maps, using a variety of different statistics. It
was pointed out that the suppression of power in the quadrupole and
octopole are aligned \cite{maxwmap}.  Further ``multipole-vector''
directions associated with these multipoles (and some other low
multipoles as well) appear to be anomalously correlated
\cite{cop04,schw04}.  There are indications of asymmetry in the power
spectrum at low multipoles in opposite hemispheres \cite{erik04a}.
Analysis of the distribution of extrema in {\it WMAP} sky maps has
indicated non-Gaussianity, and to some extent, violation of SI
\cite{lar_wan04}. The more recent WMAP maps are consistent with the
first-year maps up to a small quadrupole difference. The additional
years of data and the improvements in analysis has not significantly
altered the low multipole structures in the
maps~\cite{hin_wmap06}. Hence, `anomalies' persisted at the same
modest level of significance and are unlikely to be artifacts of
noise, systematics, or the analysis in the first year data.  The
cosmic significance of these `anomalies' remains debatable also
because of the aposteriori statistics employed to ferret them out of
the data. The WMAP team has devoted an entire publication to discuss
and present a detailed analysis of the various
anomalies~\cite{wmap7_anomal10}.

The observed CMB sky is a single realization of the underlying
correlation, hence detection of SI violation, or correlation patterns,
pose a great observational challenge. It is essential to develop a
well defined, mathematical language to quantify SI and the ability to
ascribe statistical significance to the anomalies unambiguously.  The
Bipolar spherical harmonic (BipoSH) representation of CMB correlations
has proved to be a promising avenue to characterize and quantify
violation of statistical isotropy.

Two point correlations of CMB anisotropy, $C(\hat{n}_1,\, \hat{n}_2)$,
are functions on $S^2 \times S^2$, and hence can be generally expanded
as \be
\label{bipolar} C(\hat{n}_1,\, \hat{n}_2)\, =\, \sum_{l_1,l_2,\ell,M}
\biposh{}{\ell M}{l_1 l_2} Y^{l_1l_2}_{\ell M}(\hat{n}_1,\,
\hat{n}_2)\,. \ee Here $\biposh{}{\ell M}{l_1 l_2}$ are the Bipolar
Spherical harmonic (BipoSH) coefficients of the expansion and
$Y^{l_1l_2}_{\ell M}(\hat{n}_1,\, \hat{n}_2)$ are bipolar spherical
harmonics. Bipolar spherical harmonics form an orthonormal basis on
$S^2 \times S^2$ and transform in the same manner as the spherical
harmonic function with $\ell,\, M$ with respect to
rotations. Consequently, inverse-transform of $C(\hat{n}_1,\,
\hat{n}_2)$ in eq.~(\ref{bipolar}) to obtain the BipoSH coefficients
of expansion is unambiguous.

Most importantly, the Bipolar Spherical Harmonic (BipoSH)
coefficients, $\biposh{}{\ell M}{l_1 l_2}$, are linear combinations of
{\em off-diagonal elements} of the harmonic space covariance matrix,
\be
\label{ALMvsalm} \biposh{}{\ell M}{l_1 l_2} \,=\, \sum_{m_1m_2}
\langle a_{l_1m_1}a^{*}_{l_2 m_2}\rangle (-1)^{m_2} {\mathcal C}^{\ell
M}_{l_1m_1l_2 -m_2} \ee where ${\mathcal C}_{l_1m_1l_2m_2}^{\ell M}$
are Clebsch-Gordan coefficients and completely represent the information of
the covariance matrix.  

Statistical isotropy implies that the covariance matrix is diagonal, $
\langle a_{lm}a^{*}_{l' m'}\rangle = C_{l}\,\, \delta_{ll^\prime}
\delta_{mm'}$ and hence the angular power spectra carry all
information of the field. When statistical isotropy holds BipoSH
coefficients, $\biposh{}{\ell M}{ll'}$, are zero except those with
$\ell=0, M=0$ which are equal to the angular power spectra up to a
$(-1)^l (2l+1)^{1/2}$ factor.  Therefore to test a CMB map for
statistical isotropy, one should compute the BipoSH coefficients for
the maps and look for nonzero BipoSH coefficients. {\em Statistically
significant deviations of BipoSH coefficient of map from zero would
establish violation of statistical isotropy.}

Since $A^{\ell M}_{l_1 l_2}$ form an equivalent representation of a
general two point correlation function, cosmic variance precludes
measurement of every individual $A^{\ell M}_{l_1 l_2}$. There are
several ways of combining BipoSH coefficients into different observable quantities
that serve to highlight different aspects of SI violations.  Among the
several possible combinations of BipoSH coefficients, the Bipolar
Power Spectrum (BiPS) has proved to be a useful tool with interesting
features \cite{us_apjl}. BiPS of CMB anisotropy is defined as a
convenient contraction of the BipoSH coefficients \be
\label{kappal} \kappa_\ell \,=\, \sum_{l,l',M} W_l W_{l'}\left|\biposh{}{\ell
M}{ll'}\right|^2 \geq 0 \ee where $W_l$ is the window function that
corresponds to smoothing the map in real space by symmetric kernel to
target specific regions of the multipole space and isolate the SI
violation on corresponding angular scales.

The BipoSH coefficients can be summed over $l$ and $l'$ to reduce the
cosmic variance, to as obtain reduced BipoSH (rBipoSH)
coefficients~\cite{haj_sour06} \be \label{first} A_{\ell M}=
\sum_{l=0}^{\infty}\sum_{l'=|\ell-l|}^{\ell+l} \biposh{}{\ell
M}{ll'}. \ee Reduced bipolar coefficients orientation information of
the correlation patterns. An interesting way of visualizing these
coefficients is to make a {\em Bipolar map} from $A_{\ell M}$ \be
\Theta(\hat{n}) = \sum_{\ell=0}^{\infty}\sum_{M=-\ell}^{\ell} A_{\ell
M} Y_{\ell M} (\hat{n}).  \ee The symmetry $A_{\ell M}=(-1)^M A_{\ell
-M}^*$ of reduced bipolar coefficients guarantees reality of
$\Theta(\hat{n})$.

\begin{figure}[h]
\begin{center}
\includegraphics[scale=0.7]{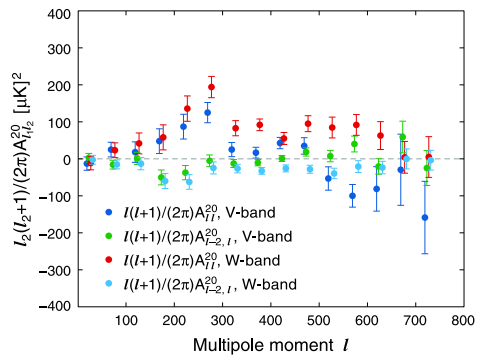}
\caption{ Figure taken from WMAP-7 yr publication on anomalies in the
CMB sky~\cite{wmap7_anomal10} shows the measured quadrupolar (bipolar
index $L=2$) bipolar power spectra for V-band and W-band WMAP data,
using the KQ75y7 mask. The spherical multipole have been binned within
uniform bands $\delta l = 50$. Only the components of the bipolar
power spectra with M = 0 in ecliptic coordinates are shown. A statistically
significant quadrupolar effect is seen, even for a single frequency
band in a single angular bin.}
\end{center}
\label{WMAP7_biposh} 
\end{figure}

It is also possible to obtain a measurable band power measure of
$A^{\ell M}_{l_1 l_2}$ coefficient by averaging $l_1$ in bands in
multipole space. Recently, the WMAP team has chosen to quantify SI
violation in the CMB anisotropy maps by the estimation $A^{\ell M}_{l
l-i}$ for small value of bipolar multipole, $L$, band averaged in
multipole $l$. Fig.~\ref{WMAP7_biposh} taken from the WMAP-7 release
paper~\cite{wmap7_anomal10} shows SI violation measured in WMAP CMB maps

CMB polarization maps over large areas of the sky have been recently
delivered by experiments in the near future. The statistical isotropy
of the CMB polarization maps will be an independent probe of the
cosmological principle.  Since CMB polarization is generated at the
surface of last scattering, violations of statistical isotropy are
pristine cosmic signatures and more difficult to attribute to the
local universe.  The Bipolar Power spectrum has been defined and
implemented for CMB polarization and show great promise~\cite{bas06}.

\section{Gravitational instability mechanism for structure formation}
\label{GI}

It is a well accepted notion that the large scale structure in the
distribution of matter in the present universe arose due to
gravitational instability from the same primordial perturbation seen
in the CMB anisotropy at the epoch of recombination. This fundamental
assumption in our understanding of structure formation has recently
found a strong direct observational
evidence~\cite{eis_sdss05,col_2Df05}.

The acoustic peaks occur because the cosmological perturbations excite
acoustic waves in the relativistic plasma of the early universe
\cite{peeb_yu70,sun_zel70,bon_efs84,bon_efs87,hol89}.  The
recombination of baryons at redshift $z\approx 1100$ effectively
decouples the baryon and photons in the plasma abruptly switching off
the wave propagation.  In the time between the excitation of the
perturbations and the epoch of recombination, modes of different
wavelength can complete different numbers of oscillation periods.
This translates the characteristic time into a characteristic length
scale and produces a harmonic series of maxima and minima in the CMB
anisotropy power spectrum.  The acoustic oscillations have a
characteristic scale known as the sound horizon, which is the comoving
distance that a sound wave could have traveled up to the epoch of
recombination.  This physical scale is determined by the expansion
history of the early universe and the baryon density that determines
the speed of acoustic waves in the baryon-photon plasma.

\begin{figure}[h]
\begin{center}
\includegraphics[scale=0.4]{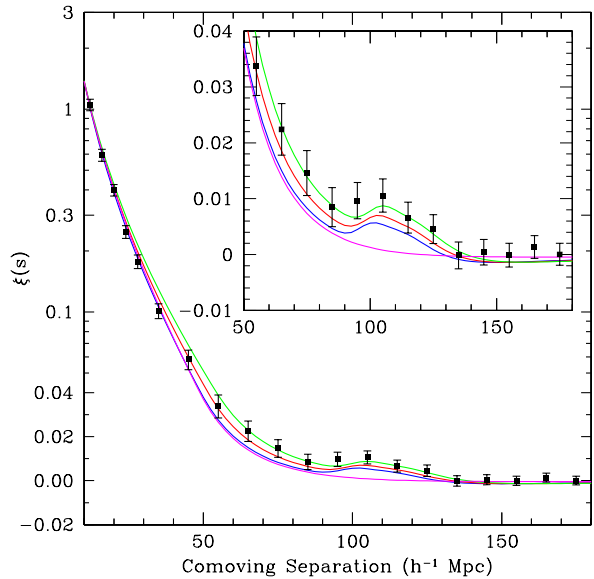}
\caption{ The large-scale redshift-space correlation function of the
SDSS LRG sample taken from Ref.~\protect{\cite{eis_sdss05}}. The
inset shows an expanded view with a linear vertical axis.  The
lower-most curve (magenta), which lacks the acoustic peak, shows a
pure CDM model ($\Omhh=0.105$).  The models are $\Omhh=0.12$
(top-most, green), 0.13 (red), and 0.14 (bottom-most with peak, blue),
all with $\Obhh=0.024$ and $n=0.98$ and with a mild non-linear
prescription folded in.  The clearly visible bump at $\sim 100\hmpc$
scale is statistically significant. }
\end{center}
\label{fig:xi} 
\end{figure}

For baryonic density comparable to that expected from Big Bang
nucleosynthesis, acoustic oscillations in the baryon-photon plasma
will also be observably imprinted onto the late-time power spectrum of
the non-relativistic matter.  This is easier understood in a real
space description of the response of the CDM and baryon-photon fluid
to metric perturbations~\cite{eis_sdss05}. An initial small
delta-function (sharp spike) adiabatic perturbation ($\delta \ln
a|_H$) at a point leads to corresponding spikes in the distribution of
cold dark matter (CDM), neutrinos, baryons and radiation (in the
`adiabatic' proportion, $1+w_i$, of the species).  The CDM
perturbation grows in place while the baryonic perturbation being
strongly coupled to radiation is carried outward in an expanding
spherical wave.  At recombination, this shell is roughly $105
h^{-1}\mpc$ in (comoving) radius when the propagation of baryons
ceases.  Afterward, the combined dark matter and baryon perturbation
seeds the formation of large-scale structure.  The remnants of the
acoustic feature in the matter correlations are weak ($10\%$ contrast
in the power spectrum) and on large scales. The acoustic oscillations
of characteristic wave-number translates to a bump (a spike softened by
gravitational clustering of baryon into the well developed dark matter
over-densities) in the correlation function at $105 h^{-1}\mpc$
separation.  The large-scale correlation function of a large
spectroscopic sample of luminous, red galaxies (LRGs) from the Sloan
Digital Sky Survey that covers $\sim 4000$ square degrees out to a
redshift of $z\sim 0.5$ with $\sim 50,000$ galaxies has allowed a
clean detection of the acoustic bump in distribution of matter in the
present universe.  Figure~\ref{fig:xi} shows the correlation function
derived from SDSS data that clearly shows the acoustic `bump' feature
at a fairly good statistical significance~\cite{eis_sdss05}. The
acoustic signatures in the large-scale clustering of galaxies provide
direct, irrefutable evidence for the theory of gravitational
clustering, notably the idea that large-scale fluctuations grow by
linear perturbation theory from $z\sim 1000$ to the present due to
gravitational instability.

\section{Primordial perturbations from Inflation}
\label{PI}

Any observational comparison based on structure formation in the
universe necessarily depends on the assumed initial conditions
describing the primordial seed perturbations.  It is well appreciated
that in `classical' big bang model the initial perturbations would
have had to be generated `acausally'. Besides resolving a number of
other problems of classical Big Bang, inflation provides a mechanism
for generating these apparently `acausally' correlated primordial
perturbations~\cite{inflpert}.

 The power in the CMB temperature anisotropy at low multipoles
($l\lsim 60$) first measured by the COBE-DMR~\cite{cobedmr} did
indicate the existence of correlated cosmological perturbations on
super Hubble-radius scales at the epoch of last scattering, except for
the (rather unlikely) possibility of all the power arising from the
integrated Sachs-Wolfe effect along the line of sight. Since the
polarization anisotropy is generated only at the last scattering
surface, the negative trough in the $C_l^{TE}$ spectrum at $l\sim 130$
(that corresponds to a scale larger than the horizon at the epoch of
last scattering) measured by WMAP first sealed this loophole, and
provides an unambiguous proof of apparently `acausal' correlations in
the cosmological perturbations~\cite{pag_wmap06,kog_wmap03,ben_wmap03}. 

Besides, the entirely theoretical motivation of the paradigm of
inflation, the assumption of Gaussian, random adiabatic scalar
perturbations with a nearly scale invariant power spectrum is arguably
also the simplest possible choice for the initial perturbations.  What
has been truly remarkable is the extent to which recent cosmological
observations have been consistent with and, in certain cases, even
vindicated the simplest set of assumptions for the initial conditions
for the (perturbed) universe discussed below.

\subsection{Nearly zero curvature of space}

The most interesting and robust constraint obtained in our quests in
the CMB sky is that on the spatial curvature of the universe. The
combination of CMB anisotropy, LSS and other observations can pin down
the universe to be flat, $\Omega_K \approx-0.02\pm 0.02$. This is
based on the basic geometrical fact that angular scale subtended in
the sky by the acoustic horizon would be different in a universe with
uniform positive (spherical), negative (hyperbolic), or, zero
(Euclidean) spatial curvature.  Inflation dilutes the curvature of the
universe to negligible values and generically predicts a (nearly)
Euclidean spatial section.

\subsection{Adiabatic primordial perturbation}

The polarization measurements provides an important test on the
adiabatic nature primordial scalar fluctuations~\footnote{ Another
independent observable is the baryon oscillation in LSS discussed in
sec~\ref{GI}}.  CMB polarization is sourced by the anisotropy of the
CMB at recombination, $z_{rec}$, the angular power spectra of
temperature and polarization are closely linked.  Peaks in the
polarization spectra are sourced by the velocity term in the same
acoustic oscillations of the baryon-photon fluid at last
scattering. Hence, a clear indication of the adiabatic initial
conditions is the compression and rarefaction peaks in the temperature
anisotropy spectrum be `out of phase' with the gradient (velocity)
driven peaks in the polarization spectra.

\begin{figure}[h]
\begin{center}
  \includegraphics[scale=0.45, angle=-90]{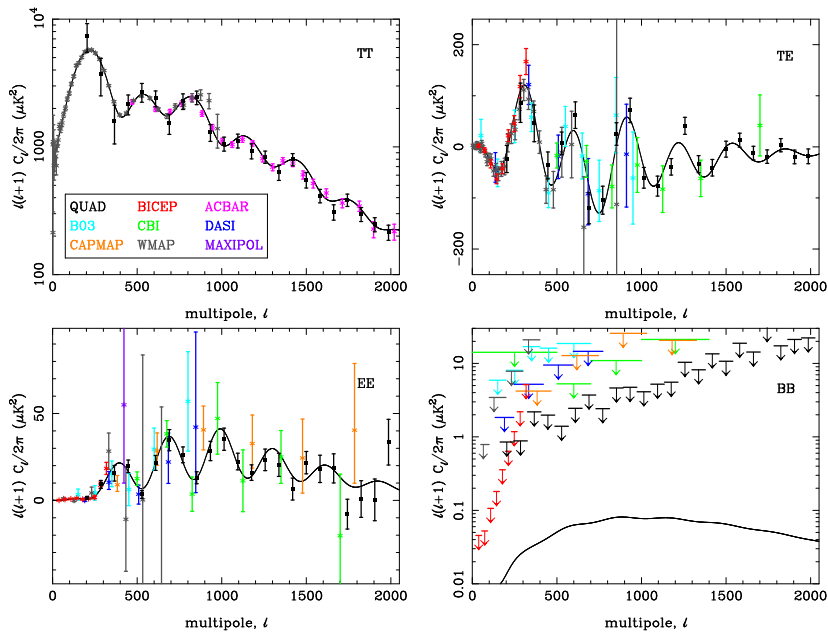}
\caption{Figure taken from Ref.~\protect{\cite{quad09}} show a
compilation of recent measurements of the angular power spectra CMB
anisotropy and polarization from a number of CMB experiments. 
The data is good enough to indicate that the peaks in EE and TE are
out of phase with that of TT as expected for adiabatic initial
conditions. The null BB detection of primary CMB signal from gravity
waves is not unexpected (given the ratio of tensor to scalar
perturbations).}
\label{cmbspecpol}
\end{center}
\end{figure}

The figure~\ref{cmbspecpol} taken from Ref.~\cite{quad09} reflects
the current observational status of CMB E-mode polarization
measurements. The recent measurements of the angular power spectrum
the E-mode of CMB polarization at large $l$ have confirmed that the
peaks in the spectra are out of phase with that of the temperature
anisotropy spectrum.

\subsection{Nearly scale-invariant power spectrum ?}
 
In a simple power law parametrization of the primordial spectrum of
density perturbation ($|\delta_k|^2 = A k^{n_s}$), the scale invariant
spectrum corresponds to $n_s=1$. Estimation of (smooth) deviations
from scale invariance favor a nearly scale invariant
spectrum~\cite{lar_wmap10}.  Current observations favor a value very close
to unity are consistent with a nearly scale invariant power spectrum.

\begin{figure}[h]
\begin{center}
  \includegraphics[scale=0.4, angle=-90]{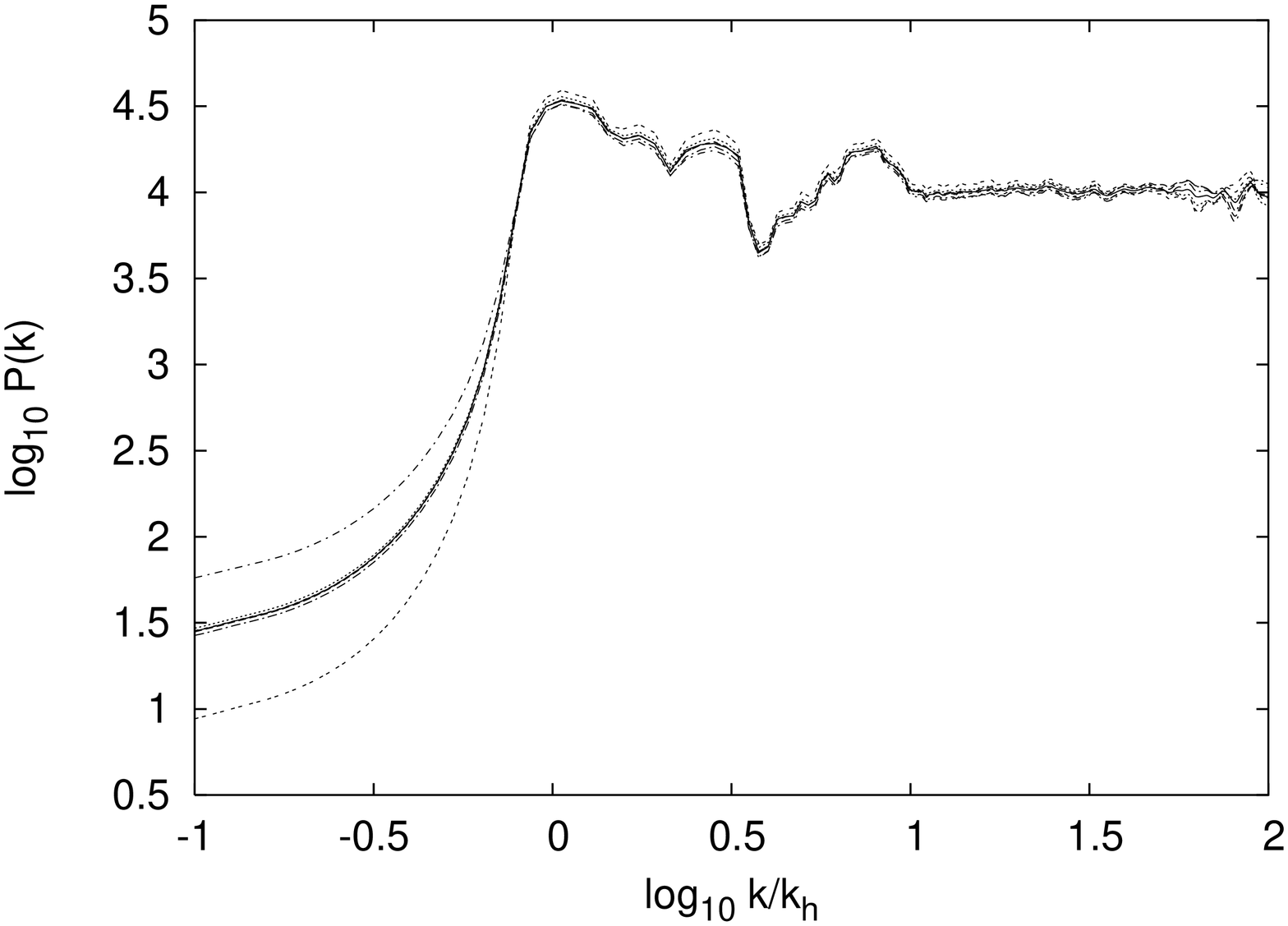}
  \caption{The primordial power spectrum recovered from the angular
  power spectrum of CMB anisotropy measured by WMAP for a concordance
  cosmological model is shown as the solid curve. Strongest deviation
  from a scale invariant Harrison-Zeldovich spectrum (which here will
  be a flat line) is the sharp infra-red cut-off at the horizon
  scale. The dotted lines correspond to the recovered spectra when
  cosmological parameters are varied within their $1-\sigma$ error
  bars and demonstrate the robustness of features in the recovered
  spectrum.}
\label{recovspec}
\end{center}
\end{figure}

While the simplest inflationary models predict that the spectral index
varies slowly with scale, inflationary models can produce strong scale
dependent fluctuations.  Many model-independent searches have also
been made to look for features in the CMB power
spectrum~\cite{bridle03,hanne04,pia03,pia05}.  Accurate measurements
of the angular power spectrum over a wide range of multipoles from the
WMAP has opened up the possibility to deconvolve the primordial power
spectrum for a given set of cosmological
parameters~\cite{max_zal02,mat_sas0203,shaf_sour04,bump05}.  The
primordial power spectrum has been deconvolved from the angular power
spectrum of CMB anisotropy measured by WMAP using an improved
implementation of the Richardson-Lucy algorithm~\cite{shaf_sour04}.
The most prominent feature of the recovered primordial power spectrum
shown in Figure~\ref{recovspec} is a sharp, infra-red cut off on the
horizon scale. It also has a localized excess just above the cut-off
which leads to great improvement of likelihood over the simple
monotonic forms of model infra-red cut-off spectra considered in the
post WMAP literature.  The form of infra-red cut-off is robust to
small changes in cosmological parameters.  Remarkably similar form of
infra-red cutoff is known to arise in very reasonable extensions and
refinement of the predictions from simple inflationary scenarios, such
as the modification to the power spectrum from a pre-inflationary
radiation dominated epoch or from a sharp change in slope of the
inflaton potential~\cite{sin_sour04}. `Punctuated Inflation' models
where a brief interruption of inflation produces features similar to
that suggested by direct deconvolution~\cite{jain_PI09}.  Wavelet
decomposition allows for clean separation of the `features' in the
recovered power spectrum on different scales~\cite{shaf_sour07}.
Recently, a frequentist analysis of the significance shows, however,
that a scale free power law spectrum is not ruled out
either~\cite{ham_shaf10}.

It is known that the assumed functional form of the primordial power
spectrum can affect the best fit parameters and their relative
confidence limits in cosmological parameter estimation. Specific
assumed form actually drives the best fit parameters into distinct
basins of likelihood in the space of cosmological parameters where the
likelihood resists improvement via modifications to the primordial
power spectrum~\cite{shaf_sour09}.  The regions where considerably
better likelihoods are obtained allowing free form primordial power
spectrum lie outside these basins. Hence, the apparently `robust'
determination of cosmological parameters under an assumed form of
$P(k)$ may be misleading and could well largely reflect the inherent
correlations in the power at different $k$ implied by the assumed form
of the primordial power spectrum. The results strongly motivate
approaches toward simultaneous estimation of the cosmological
parameters and the shape of the primordial spectrum from upcoming
cosmological data. It is equally important for theorists to keep an
open mind towards early universe scenarios that produce features in
the primordial power spectrum.

\subsection{Gaussian primordial perturbations}
\label{gauss}

The detection of primordial non-Gaussian fluctuations in the CMB would
have a profound impact on our understanding of the physics of the
early universe. The Gaussianity of the CMB anisotropy on large angular
scales directly implies Gaussian primordial
perturbations~\cite{mun95,sper_gol99} that is theoretically motivated by
inflation~\cite{inflpert}. The simplest inflationary models predict
only very mild non-Gaussianity that should be undetectable in the
WMAP data.

The CMB anisotropy maps (including the non Gaussianity analysis
carried out by the WMAP team data~\cite{lar_wmap10}) have
been found to be consistent with a Gaussian random field.  Consistent
with the predictions of simple inflationary theories, no significant
deviations from Gaussianity in the CMB maps using general tests such
as Minkowski functionals, the bispectrum, trispectrum in the three
year WMAP data~\cite{sper_wmap06,lar_wmap10}. There have however been numerous
claims of anomalies in specific forms of non-Gaussian signals in the
CMB data from WMAP at large scales (see discussion in sec.~\ref{SI}).

\subsection{Primordial tensor perturbations}

Inflationary models can produce tensor perturbations (gravitational
waves) that are predicted to evolve independently of the scalar
perturbations, with an uncorrelated power spectrum.  The amplitude of
a tensor mode falls off rapidly on sub-Hubble radius scales. The
tensor modes on the scales of Hubble-radius the line of sight to the
last scattering distort the photon propagation and generate an
additional anisotropy pattern predominantly on the largest scales. It
is common to parametrize the tensor component by the ratio $r_{k_*} =
A_{\rm t}/A_{\rm s}$, ratio of $A_{\rm t}$, the primordial power in
the transverse traceless part of the metric tensor perturbations, and
$A_{\rm s}$, the amplitude scalar perturbation at a comoving
wave-number, $k_*$ (in $\mpc^{-1}$).  For power-law models, recent
WMAP data alone puts an improved upper limit on the tensor to scalar
ratio, \ensuremath{r_{0.002} < 0.55 \mbox{ } (95\%\mbox{\ CL})} and
the combination of WMAP and the lensing-normalized SDSS galaxy survey
implies \ensuremath{r_{0.002} < 0.28 \mbox{ } (95\%\mbox{\ CL})}
~\cite{boom_polar}.

\begin{figure}[h]
\begin{center}
\includegraphics[scale=0.45]{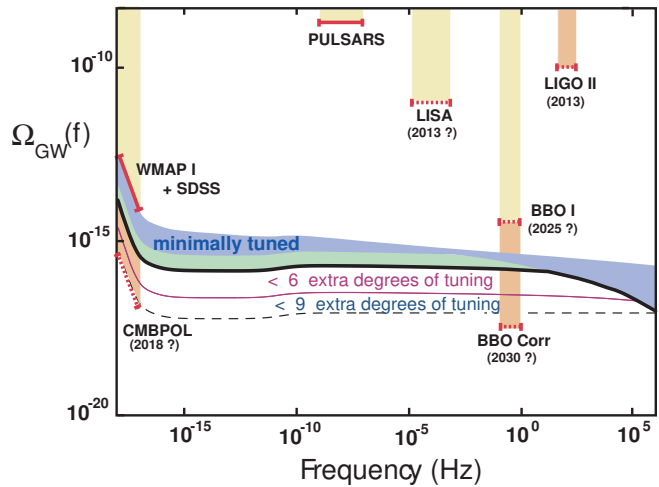} 
\caption{The figure taken from Ref.~\protect{\cite{boy_stein06}} 
shows the theoretical predictions and observational constraints on
primordial gravitational waves from inflation. The gravitational wave
energy density per logarithmic frequency interval, (in units of the
critical density) is plotted versus frequency. The shaded (blue) band
labeled `minimally tuned' represents the range predicted for simple
inflation models with the minimal number of parameters and
tunings. The dashed curves have lower values of tensor contribution,
$r$, that is possible with more fine tuned inflationary scenarios.
The currently existing experimental constraints shown are due to: big
bang nucleosynthesis (BBN), binary pulsars, and WMAP-1 (first year)
with SDSS. Also shown are the projections for LIGO (both LIGO-I, after
one year running, and LIGO-II); LISA; and BBO (both initial
sensitivity, BBO-I, and after cross-correlating receivers,
BBO-Corr). Also seen the projected sensitivity of a future space
mission for CMB polarization (CMBPol).}
\label{SGWBspec} 
\end{center}
\end{figure}

On large angular scales, the curl component of CMB polarization is a
unique signature of tensor perturbations.  Hence, the CMB
B-polarization is a direct probe of the energy scale of early universe
physics that generate the primordial metric perturbations (scalar \&
tensor). The relative amplitude of tensor to scalar perturbations,
$r$, sets the energy scale for inflation $\EI = 3.4\times
10^{16}$~GeV~$r^{1/4}$.  A measurement of $B$--mode polarization on
large scales would give us this amplitude, and hence {\em a direct
determination of the energy scale of inflation.}  Besides being a
generic prediction of inflation, the cosmological gravity wave
background from inflation would be a fundamental test of GR on cosmic
scales and the semi--classical behavior of gravity.
Figure~\ref{SGWBspec} summarizes the current theoretical
understanding, observational constraints and future possibilities for
the stochastic gravity wave background from Inflation.

\section{Conclusions}

The past few years has seen the emergence of a `concordant'
cosmological model that is consistent both with observational
constraints from the background evolution of the universe as well that
from the formation of large sale structures.  It is certainly fair to
say that the present edifice of the `standard' cosmological models is
robust. A set of foundation and pillars of cosmology have emerged and
are each supported by a number of distinct observations~\cite{me_jpo}.

The community is now looking beyond the estimation of parameters of a
working `standard' model of cosmology. There is increasing effort
towards establishing the basic principles and assumptions.  The
feasibility and promise of this ambitious goal is based on the grand
success in the recent years in pinpointing a `standard' model.  The up
coming results from the Planck space mission will radically improve
the CMB polarization measurements. There are already proposals for the
next generation dedicated satellite mission in 2020 for CMB
polarization measurements at best achievable sensitivity.

\section*{Acknowledgments}

I would like to thank the organizers for arranging an excellent
scientific meeting at a lovely location. It is a pleasure to thank and
acknowledge the contributions of students and collaborators who have
been involved with the cosmological quests in the CMB sky at IUCAA.

\end{document}